%% file: arxiv.tex
\documentclass[10pt,conference]{IEEEtran}

\usepackage{url}

\usepackage[utf8]{inputenc}

\usepackage[pdftex]{graphicx}

\usepackage[english]{babel}

\usepackage[cmex10]{amsmath}
\usepackage{amssymb}

\newcommand{\hairspace}{\hspace{1pt}}
\newcommand{\eg}{\mbox{e.\hairspace{}g}.\ }  
\newcommand{\ie}{\mbox{i.\hairspace{}e}.\ }  
\newcommand{\cf}{\mbox{c.\hairspace{}f.}\ }
\newcommand{\etal}{\mbox{et~al.}\ }

\usepackage{booktabs}

\usepackage{pifont}
\newcommand{\cmark}{\ding{51}}%
\newcommand{\xmark}{\ding{55}}%

\usepackage{mdframed} 

\usepackage{enumitem}
\setlist[enumerate]{labelindent=\parindent,topsep=1pt,itemsep=-1ex,partopsep=1ex,parsep=1.5ex}
\setlist[itemize]{labelindent=\parindent,topsep=1pt,itemsep=-1ex,partopsep=1ex,parsep=1.5ex}

\usepackage[pdftex,hyperfootnotes=false,bookmarks=false]{hyperref}
\hypersetup{
    colorlinks=true,
    linktocpage=true,
    breaklinks=true,
    bookmarksnumbered,
    bookmarksopen=true,
    bookmarksopenlevel=1,
    pdfhighlight=/O,
    pdfpagemode=UseOutlines,
    urlcolor=black,
    linkcolor=black,
    citecolor=black,
    pdfauthor={Cornelius Diekmann}
}

\usepackage{import}

\usepackage{cite}
\usepackage{nameref}

\begin{document}

\title{Demonstrating \emph{topoS}: Theorem-Prover-Based Synthesis of Secure Network Configurations}

\author{\IEEEauthorblockN{Cornelius Diekmann, Andreas Korsten, Georg Carle}\\
\IEEEauthorblockA{Technische Universit{\"a}t M{\"u}nchen \qquad \{diekmann\textbar korsten\textbar carle\}@net.in.tum.de}
}

\maketitle


\begin{abstract}
In network management, when it comes to security breaches, human error constitutes a dominant factor. 
We present our tool \emph{topoS} which automatically synthesizes low-level network configurations from high-level security goals.
The automation and a feedback loop help to prevent human errors.
Except for a last serialization step, \emph{topoS} is formally verified with Isabelle/HOL, which prevents implementation errors.
In a case study, we demonstrate \emph{topoS} by example. 
For the first time, the complete transition from high-level security goals to both firewall and SDN configurations is presented. 
\end{abstract}

\section{Introduction}
Network-level access control is a fundamental security mechanism in almost every network. 
Unfortunately, configuring network-level access control devices still is a challenging, manual, and thus error-prone task~\cite{fwviz2012,fireman2006,ZhangAlShaer2007flip}. 
It is a known and unsolved problem for over a decade that ``corporate firewalls are often enforcing poorly written rule sets''~\cite{firwallerr2004}. 
Also, ``access list conflicts dominate the misconfiguration errors made by administrators''~\cite{netsecconflicts}. 
A recent study confirms that this problem persists as a ``majority of administrators stated misconfiguration as the most common cause of failure''~\cite{sherry2012making}.
In addition, not only is implementing a policy error-prone, but also developing it is challenging, even for experienced administrators~\cite{diekmann2014verifying}.

We demonstrate our tool \emph{topoS}: a constructive, top-down greenfield approach for network security management.
\emph{topoS} translates high-level security goals to network security device configurations. 
The automatic translation steps prevent manual translation errors. 
Furthermore, \emph{topoS} visualizes the results of all translation steps to help the administrator uncover specification errors.
In addition, since all intermediate transformation steps are formally verified, the correctness of \emph{topoS} itself is guaranteed~\cite{Network_Security_Policy_Verification-AFP}. 
\emph{topoS} is built on top of recent results of the formal methods community~\cite{diekmann2014verifying,diekmann2014EPTCS}, combines these results in a novel way, and transfers the knowledge to the network management community.
The automated tool \emph{topoS} is the main technical contribution of this paper.

We first give a short overview of \emph{topoS} in Section~\ref{sec:topos}.
Then, in Section~\ref{sec:casestudy}, we present \emph{topoS} in detail with the help of a case study. 
We discuss limitations and advantages in Section~\ref{sec:discussion}, present related work in Section~\ref{sec:related}, and conclude in Section~\ref{sec:conclusion}.

\section{Overview of \emph{topoS}}
\label{sec:topos}

The security requirements of networks are usually scenario-specific.
Our tool \emph{topoS} helps to configure a network according to these needs.
It takes as input the high-level security requirements and synthesizes low-level security device configurations, \eg netfilter/iptables firewall rules or OpenFlow flow table entries.
It operates according to the following four-step process: 
\begin{enumerate}[label=\Alph*.]
	\item Formalize high-level security goals%
	\begin{enumerate}[label=\alph*.]%
		\vspace{-3\topsep}%
		\item Categorize security goals
		\item Add scenario-specific knowledge
		\item $\mathbf{\star}$ Auto-complete information
	\end{enumerate}
	\item $\mathbf{\star}$ Construct security policy
	\item $\mathbf{\star}$ Construct stateful policy
	\item $\mathbf{\star}$ Serialize security device configurations
\end{enumerate}

All steps annotated with an asterisk are supported by \emph{topoS}.
As the $\mathbf{\star}$-steps illustrate, once the security goals are specified, the process is completely automatic.
Between the automated steps, manual refinement is possible but requires re-verification.
This allows human intervention, while avoiding human error.

The automated intermediate\footnote{
We did not verify the final step (\ie serialization of security device configurations) since it is merely syntactic rewriting of the result of the previous step (\cf Sect.\ \ref{sec:statefultocongif}). 
Neither can we verify that the user expresses the security goals correctly. 
Yet, with the secure auto-completion (\cf Sect.\ \ref{sub:securitygoals}) and the visual feedback of all intermediate results, we see reason that input errors are uncovered early.} $\mathbf{\star}$-steps are proven correct for all inputs.
The proofs are verified with the interactive proof assistant \mbox{Isabelle/HOL}~\cite{isabelle2015}.  
Isabelle/\allowbreak{}HOL is an LCF-style theorem prover; the correctness of derived facts is based on the correctness of a small inference kernel. 
This architecture is very robust and widely used for over a decade. 
In general, the formal methods community treats facts machine-verified with Isabelle/HOL as well-founded truth. 
Thus, it is guaranteed that \emph{topoS} performs correct transformations~\cite{Network_Security_Policy_Verification-AFP}.
As a side note, since the transformations are proven correct once and for all for all inputs, neither has a user to prove anything manually to use \emph{topoS}, nor is Isabelle/HOL required to run \emph{topoS}.
The development of \emph{topoS} started over three years ago and features, after a large rewrite, more than $10\,\textnormal{k}$ lines of formal proof.

We will present the steps \emph{A} to \emph{D} in the following section. 
For the sake of brevity and illustrative presentation, we only present them by example. 
Mathematical background has been presented in detail previously~\cite{diekmann2014verifying,diekmann2014EPTCS,Network_Security_Policy_Verification-AFP} and in this work, we focus on its interoperability and discuss how its underlying assumptions can be fulfilled in a real-world network. 
Further details, the correctness proofs, and the interplay of the individual steps can be found in the accompanying formalization and implementation of \emph{topoS}. 

\section{\emph{topoS} by Example}
\label{sec:casestudy}
In this section, we demonstrate \emph{topoS} with a small case study. 
The scenario was chosen because it is minimal and comprehensible, but also realistic and contains many important aspects.
It runs live and is publicly available (c.f.\ Section \nameref{sec:availability}).

\begin{figure}[h!]
	\centering
  		\includegraphics[width=0.6\linewidth]{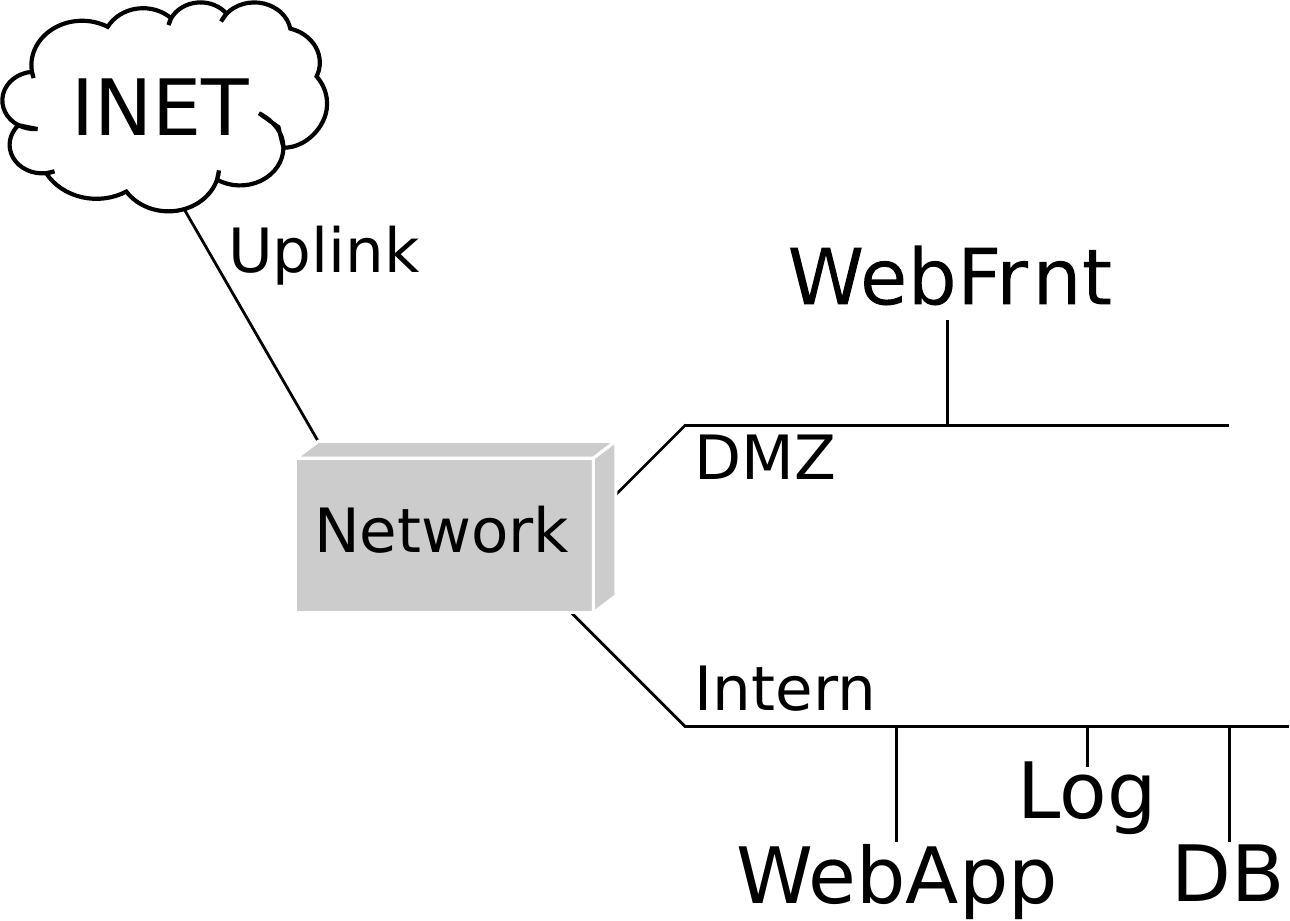}
  		\vskip-5pt
		\caption{Network Schematic}
		\label{fig:netschematic}
\end{figure}

The case study is schematically illustrated in \mbox{Fig.\ \ref{fig:netschematic}}. 
The setup hosts a news aggregation web application, accessible from the Internet (\emph{INET}). 
It consists of a web application backend server (\mbox{\emph{WebApp}}) and a frontend server (\mbox{\emph{WebFrnt}}). 
The \emph{Web\-App} is connected to a database (\emph{DB}) and actively retrieves data from the Internet. 
All servers send their logging data to a central, protected log server (\emph{Log}).

We implemented the scenario to utilize several different protocols.
A custom backend, the \emph{WebApp} was written in \texttt{python}. 
The \emph{WebFrnt} runs \texttt{lighttpd}. 
It serves static web pages directly and retrieves dynamic websites from the \emph{WebApp} via \texttt{FastCGI}. 
All components send their \texttt{syslog} messages via UDP (RFC~5426) to \emph{Log}.

\subsection{Formalizing Security Goals}
\label{sub:securitygoals}
The security goals are expressed as security invariants over the network's connectivity structure.
An invariant consists of a generic part (the semantics) and scenario-specific information. 
The generic part defines the type and general meaning. 
Our generic invariants currently defined are summarized in \mbox{Table~\ref{tab:invariantemplates}}. 

\newcommand{\ignore}[1]{#1}
\begin{table}[h!bt]
\centering
\caption{Generic Security Invariants}
\vskip-5pt
\label{tab:invariantemplates}
\begin{footnotesize}
\begin{tabular}{ @{} l @{\hspace*{0.8em}} c @{\hspace*{0.8em}} p{5.9cm} @{}}%
	\toprule
	Name          \ignore{& $\Phi$} & Description  \\
	\midrule
	Bell LaPadula \ignore{& \cmark} & Label-based Information Flow Security\\
	Comm.\ Partners \ignore{& \cmark} & Simple ACLs (Access Control Lists)  \\
	Comm.\ With \ignore{& \xmark} & White-listing transitive ACLs   \\
	Not Comm.\ With \ignore{& \xmark} & Black-listing transitive ACLs   \\
	Dependability \ignore{& \xmark} & Limit dependence on certain hosts \\
	Domain Hierarchy \ignore{& \cmark} & Hierarchical control structures \\
	Refl \ignore{& \cmark} & Allow/deny reflexive flows. Can lift symbolic policy identifiers to role names \\
	NonInterference \ignore{& \xmark} & Transitive non-interference properties \\
	Security Gateway \ignore{& \cmark} & Master/Slave relationships \\
	Sink \ignore{& \cmark} & Information sink \\
	Subnets \ignore{& \cmark} & Collaborating, protected host groups\\
 \bottomrule%
\end{tabular}%
\end{footnotesize}
\end{table}

To construct a scenario-specific invariant, a generic invariant is instantiated with scenario-specific knowledge. 
This is done by specifying host attributes~\cite{diekmann2014verifying}. 
These invariants and the list of entities (\emph{INET}, \mbox{\emph{WebApp}}, \mbox{\emph{WebFrnt}}, \emph{DB}, \emph{Log}) is the only input needed. 
For this scenario, the following four invariants are expressed, formalized in \mbox{Fig.\ \ref{fig:runexsecinvars}}.

\begin{enumerate}
\item First, as illustrated in Fig.\ \ref{fig:netschematic}, \emph{DB}, \emph{Log} and \emph{WebApp} are labeled as internal hosts. 
The \emph{WebFrnt} must be accessible from outside and is thus labeled as DMZ member. 
This is captured in the \emph{Subnets} invariant. 

\item Next, it is expressed that the logging data must not leave the log server. 
Therefore, using the \emph{Sink} invariant, \emph{Log} is classified as information sink.

\item Using the \emph{Bell LaPadula} invariant, it is specified that \emph{DB} contains confidential information. 
Since it sends its log data to the log server, this log server is also assigned the confidential security clearance. 
Finally, the \emph{Web\-App} is allowed to retrieve data from the \emph{DB} and to publish it to the \emph{WebFrnt}. 
Therefore, the \emph{Web\-App} is trusted and allowed to declassify the data.

\item Finally, an access control list specifies that only \emph{Web\-App} may access the \emph{DB}. 
\end{enumerate}


\begin{figure}[bht]
\centering
\fbox{
\begin{minipage}{0.99\linewidth}
\begin{flushleft}
\smallskip
Subnets $\lbrace \mathrm{DB} \mapsto \mathit{internal},\ \mathrm{Log} \mapsto \mathit{internal},\ \newline
\hspace*{4.21em} \mathrm{WebApp} \mapsto \mathit{internal},\ \mathrm{WebFrnt} \mapsto \mathit{DMZ}
\rbrace$

\medskip

Sink $\lbrace \mathrm{Log} \mapsto \mathit{Sink} \rbrace$

\medskip

Bell LaPadula $\lbrace \mathrm{DB} \mapsto \mathit{confidential},\ \mathrm{Log} \mapsto \mathit{confidential},\ \newline
\hspace*{7.0em} \mathrm{WebApp} \mapsto \mathit{declassify} \ \mathit{(trusted)} \rbrace$

\medskip

Comm.\ Partners $\lbrace
\mathrm{DB} \mapsto \mathit{Access\ allowed\ by}: \mathrm{WebApp}
\rbrace$\\
\smallskip
\end{flushleft}
\end{minipage}
}
\vskip-5pt
\caption{Security Invariants (Case Study)}
\label{fig:runexsecinvars}
\end{figure}

In this example, several hosts do not have attributes assigned for all invariants. 
It is sufficient to supply an incomplete host attribute specification, since they are automatically and securely completed by \emph{topoS}. 
Previous work~\cite{diekmann2014verifying} discusses the details.\footnote{The security of the auto-completion is guaranteed w.r.t.\ the provided information, \ie the auto-completion can never lead to an unnoticed security problem, given enough information is provided. 
For example, information-leakage is always uncovered, given all confidential data sources are specified.
However, if an administrator forgets to label a confidential data source, information leakage can occur.
It is trivially possible to design explicit whitelisting invariants which auto-complete to some `\emph{deny}' property. 
On the downside, this requires lots of manual configuration effort, which is avoided by the invariants utilized in this paper.
Roughly speaking the auto-completion fulfills: ``\textit{the more information provided, the more secure the whole system}''.}
Once the invariants are specified, their management scales well in the face of changes:
When a new host is added to the network, issues are handled by the auto-completion: 
either, the new host causes a violation, which is consequently uncovered, or it can be added without any further changes.
Invariants are composable and modular by design, helping structured representation and archiving of knowledge. 
In the worst case, inconsistent security invariants may be specified accidentally. 
This only results in an overly strict security policy being computed, which can be identified in the following step.

It has been shown that a special class of invariants, called $\Phi$-struc\-tured, exhibits several nice mathematical properties~\cite{diekmann2014verifying} (\cf Tab.\ \ref{tab:invariantemplates}). 
A $\Phi$-struc\-tured invariant asserts a predicate for every policy rule.  
This predicate must only depend on the sender, receiver, and their host attributes. 
In particular, these invariants and their derived algorithms are very efficiently computable. 
It is also due to the $\Phi$-struc\-tured invariants that a maximum-permissive security policy is uniquely defined.

\subsection{Constructing the Security Policy}
A network's end-to-end connectivity structure, \ie a global access control matrix, corresponds to the \emph{security policy}. 
Here, we utilize the textbook definition that a policy consists of the \emph{rules} which ensure that the network is in a secure state. 
In contrast, the security goals are expressed as \emph{invariants} over the policy and reside on an higher abstraction level.

\begin{figure}[htb]
	\centering
	\begin{minipage}{0.48\linewidth}\centering
  		\includegraphics[width=0.99\textwidth]{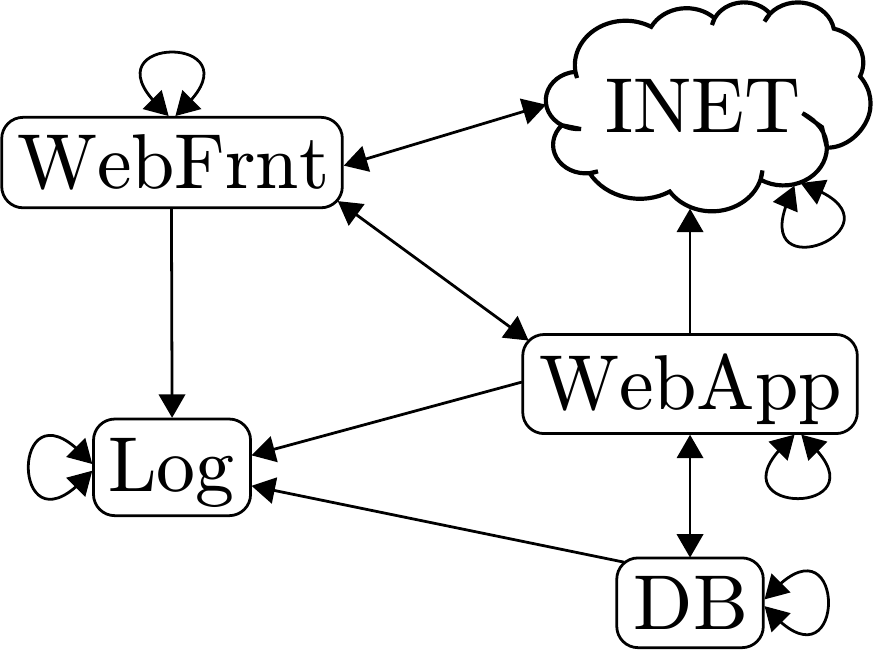}
  		\vskip-5pt
		\caption[Security Policy computed]{Security Policy\\ \phantom{\mbox{\quad (computed)}}} 
		\label{fig:secpol}
	\end{minipage}
	\hspace*{\fill}
	\begin{minipage}{0.48\linewidth}\centering
  		\includegraphics[width=0.99\textwidth]{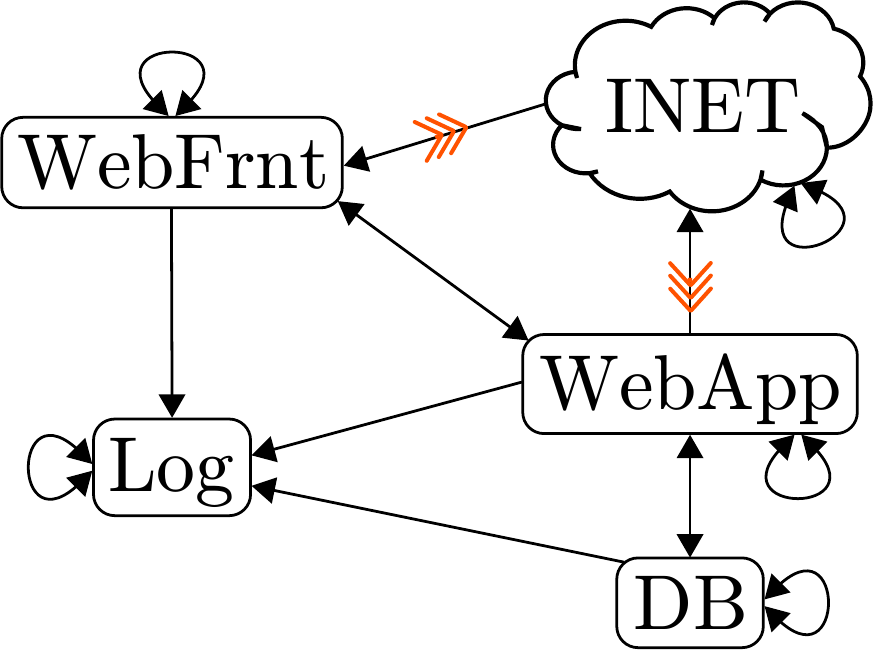}
  		\vskip-5pt
		\caption[Stateful Policy]{Stateful Policy\\ \phantom{\mbox{\quad (computed)}}}
		\label{fig:statefulpol}
	\end{minipage}
\end{figure}

Graphically, a policy can be illustrated as a directed graph. 
The case study's policy, illustrated in Fig.\ \ref{fig:secpol}, was automatically computed from the security invariants.

The algorithm to transform a set of security invariants into a policy starts with the allow-all policy and iteratively removes undesired rules.
This is always possible if (and only if~\cite{diekmann2014verifying}, Theorem\hairspace{}\hairspace{}1) the invariants hold for the deny-all policy; a static requirement which is only to be proven once for a generic invariant. 
The algorithm is sound. 
It is also complete for the invariants utilized in this example (and for $\Phi$-struc\-tured invariants in general~\cite{Network_Security_Policy_Verification-AFP}). 

In our example, the administrator decides to manually refine the policy: there is no need for the web frontend to connect to the Internet. 
Therefore, this flow is prohibited. 
After this manual refinement, the security invariants are re-verified.

\subsection{Constructing the Stateful Policy}
\label{sec:statefulpolicy}
The derived policy may appear adequate from a theoretical point of view but has one major problem when it comes to implementation:
The \emph{WebApp} can connect to the Internet, but the policy does not specify whether the Internet may answer this request (same for the \emph{WebFrnt} after manual refinement). 
Obviously, for this scenario, answers should be permitted; otherwise, no one would be able to use the service. 
In contrast, the Log server uses the \texttt{syslog} protocol over UDP (RFC~5426).  
This protocol uses a unidirectional UDP channel and it is explicitly specified for security reasons that this is the only way the communication with the log server is permitted.

Therefore, it must be distinguished between stateful and purely unidirectional rules. 
We extend the security policy to additionally specify whether a flow might be stateful (\ie answers to requests are allowed).
Note that a flow with the stateful attribute might allow packets in the opposite direction of the policy rule and thus potentially violate security invariants.
Defining the following two consistency criteria, the stateful attributes can be computed automatically~\cite{diekmann2014EPTCS}:
\begin{enumerate}
	\item No information flow violation must occur
	\item No access control side effects must be introduced
\end{enumerate}
To compute the stateful policy, not only a single rule but a set $S$ is to be upgraded to stateful rules. 
However, the interaction of the rules and answer paths of $S$ must not introduce negative implications. 
Therefore, in particular to verify lack of side effects, all security policies derived from upgrading all \emph{subsets} of $S$ must be verified. 
A naive approach would require exponential complexity. 
We proved that this can be done more efficiently, particularly in linear time for $\Phi$-structured invariants~\cite{diekmann2014EPTCS}. 
This insight provides an algorithm for computing the stateful policy from the security policy and the invariants. 
It is proven sound \cite[Theorem\hairspace{}\hairspace{}2]{diekmann2014EPTCS} and complete w.r.t.\ the two criteria individually~\cite{diekmann2014EPTCS}. 
Multiple solutions for a stateful policy may exist; a user may set preferences.

For the case study, this results in a policy where the Internet can set up connections to the web frontend, likewise, the web backend can set up connections to the Internet. 
However, the logging channels are purely unidirectional UDP (stateful connections would introduce an information flow violation).
We will call this the stateful policy.
It is illustrated in Fig.\ \ref{fig:statefulpol}.

\subsection{Serializing Security Device Configurations} 
\label{sec:statefultocongif}
Till now, the network of Fig.\ \ref{fig:netschematic} was considered a black box.
In this section, the stateful policy is serialized to configurations for real network security devices.
Though the serialization step is merely syntactic rewriting of the stateful policy, care must be taken to correctly transfer the semantics. 
\emph{topoS} must fulfill the following three assumptions. 
\begin{description}
	\item[Structure] The enforced network connectivity structure must exactly coincide with the policy. This requires that the links are confidential and integrity protected. 
	\item[Authenticity] The policy's entities must match their network representation (\eg IP/MAC addresses). In particular, no impersonation or spoofing must be possible. 
	\item[State] The stateful connection handling must match the stateful policy's semantics.
\end{description}

One policy entity may correspond to several entities in the network. 
For example, deployed with load-balancing, \emph{Web\-App} corresponds a set of backend servers. 
In such cases, special care must be taken for reflexive policy rules.
For the sake of brevity, we only present a one-to-one mapping between policy entities and their network representatives in this paper.

\noindent
We present two possibilities to implement the policy. 

\subsubsection{Firewall \& Central VPN Server}
\label{deploy:openvpn}
\begin{figure*}[ht!bp]
\fbox{
\begin{minipage}{0.99\linewidth}
\begin{flushleft}
\begin{small}
\texttt{\noindent
FORWARD DROP\\
-A FORWARD -i tun0 -s $\mathit{\$WebFrnt\_ipv4}$ -o tun0 -d $\mathit{\$Log\_ipv4}$ -j ACCEPT \\
-A FORWARD -i tun0 -s $\mathit{\$WebFrnt\_ipv4}$ -o tun0 -d $\mathit{\$WebApp\_ipv4}$ -j ACCEPT \\
-A FORWARD -i tun0 -s $\mathit{\$DB\_ipv4}$ -o tun0 -d $\mathit{\$Log\_ipv4}$ -j ACCEPT \\
-A FORWARD -i tun0 -s $\mathit{\$DB\_ipv4}$ -o tun0 -d $\mathit{\$WebApp\_ipv4}$ -j ACCEPT \\
-A FORWARD -i tun0 -s $\mathit{\$WebApp\_ipv4}$ -o tun0 -d $\mathit{\$WebFrnt\_ipv4}$ -j ACCEPT \\
-A FORWARD -i tun0 -s $\mathit{\$WebApp\_ipv4}$ -o tun0 -d $\mathit{\$DB\_ipv4}$ -j ACCEPT \\
-A FORWARD -i tun0 -s $\mathit{\$WebApp\_ipv4}$ -o tun0 -d $\mathit{\$Log\_ipv4}$ -j ACCEPT \\
-A FORWARD -i tun0 -s $\mathit{\$WebApp\_ipv4}$ -o eth0 -d $\mathit{\$INET\_ipv4}$ -j ACCEPT \\
-A FORWARD -i eth0 -s $\mathit{\$INET\_ipv4}$ -o tun0 -d $\mathit{\$WebFrnt\_ipv4}$ -j ACCEPT \\
-I FORWARD -m state --state ESTABLISHED -i eth0 -s $\mathit{\$INET\_ipv4}$ 
 -o tun0 -d $\mathit{\$WebApp\_ipv4}$  -j ACCEPT\\ %
-I FORWARD -m state --state ESTABLISHED -i tun0 -s $\mathit{\$WebFrnt\_ipv4}$ 
 -o eth0 -d $\mathit{\$INET\_ipv4}$  -j ACCEPT\\ %
}
\end{small}
\end{flushleft}
\end{minipage}
}
\vskip-5pt
\caption{VPN Server Firewall Rules (can be loaded with \texttt{iptables})}
\label{fig:vpnfirewall}
\end{figure*}

\begin{figure*}[ht!bp]
\fbox{
\begin{minipage}{0.99\linewidth}
\begin{flushleft}
\begin{small}
\texttt{\noindent
\# ARP Request\\
  in\_port=$\mathit{\$port\_src}$ dl\_src=$\mathit{\$mac\_src}$ dl\_dst=ff:ff:ff:ff:ff:ff 
  arp arp\_sha=$\mathit{\$mac\_src}$ \hfill $\hookleftarrow$\\ 
  \qquad arp\_spa=$\mathit{\$ip4\_src}$ arp\_tpa=$\mathit{\$ip4\_dst}$ 
  priority=40000 action=mod\_dl\_dst:$\mathit{\$mac\_dst}$,output:$\mathit{\$port\_dst}$\\
\medskip
\# ARP Reply\\
  dl\_src=$\mathit{\$mac\_dst}$ dl\_dst=$\mathit{\$mac\_src}$ 
  arp arp\_sha=$\mathit{\$mac\_dst}$ arp\_spa=$\mathit{\$ip4\_dst}$ arp\_tpa=$\mathit{\$ip4\_src}$ \hfill $\hookleftarrow$\\
  \qquad priority=40000 action=output:$\mathit{\$port\_src}$\\
\medskip
\# IPv4 one-way\\
  in\_port=$\mathit{\$port\_src}$ dl\_src=$\mathit{\$mac\_src}$ ip nw\_src=$\mathit{\$ip4\_src}$ nw\_dst=$\mathit{\$ip4\_dst}$ 
   priority=40000 \hfill $\hookleftarrow$\\\qquad action=mod\_dl\_dst:$\mathit{\$mac\_dst}$,output:$\mathit{\$port\_dst}$\\
\medskip
\# if $\mathit{src}$ (res.\ $\mathit{dst}$) is INET, replace $\mathit{\$ip4\_src}$ (resp.\ $\mathit{\$ip4\_dst}$) with * and decrease the priority
}%
\end{small}
\end{flushleft}%
\end{minipage}
}
\vskip-5pt
\caption{OpenFlow Flow Table Template (can be loaded with \texttt{ovs-vsctl set-fail-mode $\mathit{\$switch}$ secure \&\& ovs-ofctl add-flows})}%
\label{fig:openflowrules}
\end{figure*}

All entities connect to a central OpenVPN server which enforces the policy. 
Entities are bound to their policy name with \mbox{X.509} certificates. 
Every entity sets up a layer~3 (\texttt{tun}) VPN connection with the server.
The server authenticates entities by their certificate and centrally assigns IP addresses. 
IP spoofing over the tunnel is prevented. 
This provides authenticity~(\cmark).
Firewalling is applied at the server; the stateful policy is directly translated to \texttt{iptables} rules, shown in \mbox{Fig.\ \ref{fig:vpnfirewall}}. 
With this, the stateful semantics (\cmark) and structure (\cmark) are enforced.

\subsubsection{SDN}
With complete control over the network, as is the case with data centers, a Software-Defined Network (SDN) may be used to implement the policy. 
Usually, a data center is a flat layer~2 network~\cite{bari2013data} and we need to contain layer~2 broadcasting and attacks. 
For this, an entity's switch port must be known. 
We install OpenFlow rules which prevent MAC, IP, and ARP spoofing. 
Figure~\ref{fig:openflowrules} illustrates a template for generating a stateless rule from $\mathit{src}$ to $\mathit{dst}$. 
The first rule allows ARP requests.
Note that we rewrite the layer~2 broadcast addresses directly to the immediate receiver's address.
Rule two allows the ARP responses.
Both rules ensure that only valid ARP queries and responses are sent and received in the network.\footnote{
  For the sake of simplicity, this implementation is designed such that it gets along without an SDN controller. 
  This introduces a small hidden information flow channel (structure \xmark): the ARP responses.
  For example in Fig.\ \ref{fig:statefulpol}, \emph{Log} may use a timing channel or the ARP \texttt{OPER} field to exfiltrate information. 
  However, the side-channel is easily removed when an SDN controller answers all ARP requests (structure \cmark); all necessary information is present.
  }
The third rule allows IPv4 traffic.
For stateful rules, the opposite direction of \mbox{Fig.\ \ref{fig:openflowrules}}, \ie $\mathit{src}$ and $\mathit{dst}$ swapped, is added.
Any unmatched packets are dropped. 
With this set of rules, a mapping of policy identifiers to MAC and IP addresses is enforced. 
Also, correct address resolution is enforced (authenticity~\cmark). 
Without the ARP information leak, the desired connectivity structure (\cmark) is enforced.
The setup does not provide stateful handling (\xmark) by default. 
However, a network firewall or SDN firewall app can provide the desired state (\cmark) handling.

\section{Discussion}
\label{sec:discussion}
\subsection{Limitations}
The process supported by \emph{topoS} currently has two main limitations. 
First, it is completely static. 
For example, the mapping of policy entity names to their network representatives is done statically and manually. 
In general, \emph{naming} is a complex (but orthogonal) issue. 
This information should usually be managed by a resource and account management system or directory service.
Second, only one security device as backend is currently supported. 
However, related work suggests that this gap can be easily bridged, \eg by translating to a one big switch abstraction~\cite{monsanto2013composingonebigswitch,icfp2015smolkanetkatcompiler}.
Many networks additionally employ a variety of heterogeneous, vendor-specific middleboxes. 
In future work, it might be worth investigating to which extent additional low-level device features (\eg DHCP, IPv6, timeouts for stateful rules, ...) should be configurable on each abstraction layer.

\subsection{Advantages}
The presented process provides three novel advantages. 
First, it \emph{bridges several abstraction levels} in a uniform way. 
The intermediate results are well-specified, which allows manual intervention, visualization, and adding features.
Second, the theoretical background is \emph{completely formally verified}. 
Thus, \emph{topoS} is more than an academic prototype but a highly trustworthy tool. 
In addition, \emph{topoS}'s library can be reused, extended, exported to several languages, and adapted to fit the needs of other frameworks.
Finally, with the formal background, \emph{topoS} is a first step towards high-assurance certification. 

Third, \emph{topoS} can scale to large networks w.r.t.\ theory~(\emph{i}), computational complexity~(\emph{ii}), and management complexity~(\emph{iii}). 
The theoretical foundation (\emph{i}) scales to arbitrary networks. 
The computational complexity (\emph{ii}) depends on the type of security invariants. 
New invariants with arbitrary computational complexity can be developed for \emph{topoS}. 
However, we found that usually only $\Phi$-structured invariants are needed, which implies the following computational complexity: $\mathrm{O}(\vert \mathit{invariants} \vert \cdot \vert \mathit{entities} \vert^2)$. 
An evaluation of \emph{topoS}'s most expensive step has been presented previously~\cite{diekmann2014verifying}.
Finally, (\emph{iii}) the complexity of managing an invariant (with exception for the ACL invariants) is linear in the number entities. 
Due to the auto-completion, it is actually better than linear.
Thus, the management complexity of \emph{topoS} is roughly linear in the number of invariants and entities.

\section{Related Work}
\label{sec:related}
To discuss related work, we first define four management abstraction layers to subsequently classify related work.

\begin{description}
	\item[Security Invariants]
		Defines the high-level security goals.
		Representable as predicates.
		For example, \mbox{Fig.\ \ref{fig:runexsecinvars}}.
	\item[Access Control Abstraction]
		Defines the allowed accesses between policy entities.
		Representable as access control matrix.
		For example, \mbox{Fig.\ \ref{fig:secpol}}.
	\item[Interface Abstraction]
		Defines a model of the complete network topology.
		Representable as a graph, packets are forwarded between the entity's interfaces.
	\item[Box Semantics]
		Describes the semantics (\ie behavior) of individual network boxes.
		Usually, the semantics are vendor-specific (\eg iptables, Cisco ACLs, Snort IDS, ...). 
\end{description}

\begin{figure}[h!tb]
	\centering
	\vskip5pt
		\resizebox{0.99\linewidth}{!}{
		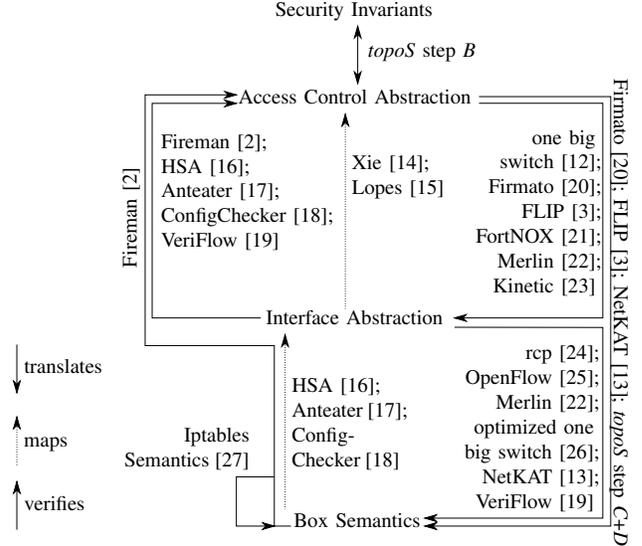}
  		\vskip-5pt
		\caption{Four Layer Abstraction in Related Work}
		\label{fig:relatedworkabstractionlayers}
\end{figure}

In \mbox{Fig.\ \ref{fig:relatedworkabstractionlayers}}, we summarize how related work bridges the abstraction layers.
Related work may bridge these layers vertically or work horizontally on artifacts at one layer.
A direct arrow from the Access Control Abstraction to the Box Semantics (and vice versa) means that the solution only applies to a single enforcement box. Solutions such as Firmato and Fireman achieve more and are thus listed multiple times.

%
%
\newcommand{\boxSemanticsToAccessControlAbstraction}[0]{Fireman~\cite{fireman2006}}
\newcommand{\InterfaceAbstractionToAccessControl}[0]{Fireman~\cite{fireman2006}; HSA~\cite{kazemian2012HSA}; Anteater~\cite{Mai2011anteater}; Config\-Checker~\cite{alshaer2009configchecker}; VeriFlow~\cite{khurshid2013veriflow}} 
\newcommand{\InterfaceAbstractionMAPSAccessControl}[0]{Xie~\cite{xie2005static}; Lopes~\cite{lopes2013msrnetworkverificationprogram}}
\newcommand{\BoxSemanticsMAPSInterfaceAbstraction}[0]{HSA~\cite{kazemian2012HSA}; Anteater~\cite{Mai2011anteater}; Config\-Checker~\cite{alshaer2009configchecker}}
\newcommand{\AccessControlToInterfaceAbstraction}[0]{one big\phantom{;} switch~\cite{monsanto2013composingonebigswitch}; Firmato~\cite{bartal1999firmato}; FLIP~\cite{ZhangAlShaer2007flip}; FortNOX~\cite{Porras2012FortNOX}; Merlin~\cite{soule2014merlin}; Kinetic~\cite{kinetic2015}\phantom{;}}
\newcommand{\AccessControlBISinvars}[0]{\emph{topoS} \mbox{step \emph{B}}} 
\newcommand{\AccessControlToBoxSemantics}[0]{Firmato~\cite{bartal1999firmato}; FLIP~\cite{ZhangAlShaer2007flip}; NetKAT~\cite{icfp2015smolkanetkatcompiler}; \emph{topoS} step \emph{C}+\emph{D}}
\newcommand{\InterfaceAbstractionToBoxSemantics}[0]{rcp~\cite{Caesar2005rcp}; OpenFlow~\cite{mckeown2008openflow}; Merlin~\cite{soule2014merlin}; optimized one\phantom{;} \mbox{big switch}~\cite{Kang2013onebigswitchabstraction}; NetKAT~\cite{icfp2015smolkanetkatcompiler}; VeriFlow~\cite{khurshid2013veriflow}\phantom{;}}
\newcommand{\BoxSemantics}[0]{Iptables \mbox{Semantics}~\cite{diekmann2015fm}}
%

\begin{figure}[h!tb]
	\centering
  		\includegraphics[width=0.8\linewidth]{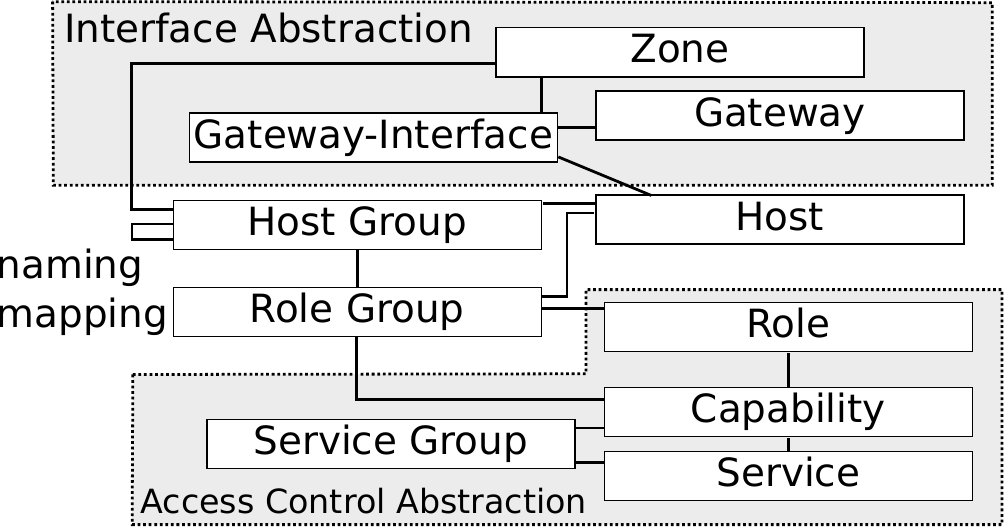}
  		\vskip-5pt
		\caption{Firmato ERM}
		\label{fig:firmatoerm}
\end{figure}

Firmato~\cite{bartal1999firmato} is the work closest related to \emph{topoS}.
It defines an entity relationship model 
to structure network management and compile firewall rules from it, illustrated in \mbox{Fig.\,\ref{fig:firmatoerm}}. 
Firmato focuses on roles, which correspond to policy entities in our model.
A role has positive capabilities and is related to other roles, which can be used to derive an access control matrix.
Zones, Gateway-Interfaces and Gateways define the network topology, which corresponds to the interface abstraction.
As illustrated in \mbox{Fig.\ \ref{fig:firmatoerm}}, the abstraction layers identified in this work can also be identified in Firmato's model.
The Host Groups, Role Groups and Hosts definitions provide a mapping from policy entities to network entities, which is Firmato's approach to the naming problem.
Similar to Firmato (with more support for negative capabilities) is FLIP~\cite{ZhangAlShaer2007flip}, which is a high-level language with focus on \emph{service} management (\eg allow/deny HTTP). 
Essentially, both FLIP and Firmato enhance the Access Control Matrix horizontally by including layer four port management and traverse it vertically by serializing to firewall rules.

As illustrated in \mbox{Fig.\ \ref{fig:relatedworkabstractionlayers}},
Fireman~\cite{fireman2006} is a counterpart to Firmato.
It verifies firewall rules against a global access policy.
In addition, Fireman provides verification on the same horizontal layer (\ie finding shadowed rules or inter-firewall conflicts, which do not affect the resulting end-to-end connectivity but are still most likely an implementation error).
Abstracting to its uses, one may call rcc~\cite{Feamster2005rcc} the fireman for BGP.

Header Space Analysis \allowbreak{}\mbox{(HSA)~\cite{kazemian2012HSA}}, Anteater \cite{Mai2011anteater}, and Config\-Checker~\cite{alshaer2009configchecker} verify several horizontal safety properties on the interface abstraction, such as absence of forwarding loops.
By analyzing reachability~\cite{xie2005static,lopes2013msrnetworkverificationprogram,Mai2011anteater,alshaer2009configchecker,kazemian2012HSA}, horizontal consistency of the interface abstraction with an access control matrix can also be verified. 
Verification of incremental changes to the interface abstraction can be done in real-time with VeriFlow~\cite{khurshid2013veriflow}, which can also prevent installation of violating rules. 
These models of the interface abstraction have many commonalities: 
The network boxes in all models are stateless and the network topology is a graph, connecting the entity's interfaces.
A function models packet traversal at a network box. 
These models could be considered as a giant (extended) finite state machine (FSM), where the state of a packet is an (interface$\times$packet) pair and the network topology and forwarding function represent the state transition function~\cite{lopes2013msrnetworkverificationprogram,zhang2012erificationswitching}.
Anteater~\cite{Mai2011anteater} differs in that interface information is implicit and packet modification is represented by relations over packet histories.

Most analysis tools make simplifying assumptions about the underlying network boxes.
Diekmann \etal\cite{diekmann2015fm} present simplification of iptables firewalls to make complex real-world firewalls available for tools with simplifying assumptions.

NetKAT \cite{icfp2015smolkanetkatcompiler} is a SDN programming language with well-defined semantics. 
It features an efficient compiler for local, global, and virtual programs to flow table entries.

Craven \etal\cite{craven2011policyrefinement} present a generalized (not network-specific) process to translate access control policies, enhanced with several aspects, to enforceable device-specific policies; the implementation requires a model repository of box semantics and their interplay.
Pahl delivers a data-centric, network-specific approach for managing and implementing such a repository, further focusing on things~\cite{Pahl2015IM}. 

FortNOX~\cite{Porras2012FortNOX} horizontally enhances the access control abstraction as it assures that rules by security apps are not overwritten by other apps.
Technically, it hooks up at the access control/interface abstraction translation. 
Kinetic~\cite{kinetic2015} is an SDN language which lifts static policies (as constructed by \emph{topoS}) to dynamic policies.
To accomplish this, an administrator can define a simple FSM which dynamically (triggered by network events) switches between static policies. 
In addition, the FSM can be verified with a model checker.
Features are horizontally added to the interface abstraction:
a routing policy allows specifying \emph{paths} of network traffic~\cite{Kang2013onebigswitchabstraction}. 
Merlin~\cite{soule2014merlin} additionally supports bandwidth assignments and network function chaining.
Both translate from a global policy to local enforcement and Merlin provides a feature-rich language for interface abstraction policies.

\section{Conclusion}
\label{sec:conclusion}
We presented \emph{topoS}, a fully verified tool to manage network-level access control.
It was demonstrated by example; nevertheless, the correctness proofs are universally valid and \emph{topoS} is applicable to any larger network.
The example demonstrates that a traditional network segmentation into \emph{internal} and \emph{DMZ} cannot cope with complex security goals and the traditional thought model of structure by IP ranges is no longer appropriate. 
In contrast, \emph{topoS} only requires the high-level security goals and can automatically translate them to low-level configurations, such as firewall rules or SDN flow table entries.
During the translation, all intermediate results are well-defined, accessible, and can be visualized.
This provides feedback and allows manual refinement of them, including manual optimizations on lower abstraction layers.
After manual refinement, re-verification is run to avoid human error.
For the first time, the complete, automated, and verified transition from high-level security goals to both Firewall and SDN configurations was presented.

\section*{Availability \& Acknowledgements}
\label{sec:availability}
\noindent
Our tool \emph{topoS} and the correctness proofs can be obtained at
\begin{center}\vskip-0.8em
\url{https://github.com/diekmann/topoS/} \ \ or \ \ \cite{Network_Security_Policy_Verification-AFP}
\end{center}\vskip-0.8em
The formalization of the case study is \url{Distributed_WebApp.thy}
It runs live at: 
\url{http://otoro.net.in.tum.de/goals2config/}

This work has been supported by the German Federal Ministry of
Education, EUREKA project SASER, grant 16BP12304, and project SURF, grant 16KIS0145, and by the European Commission, project SafeCloud, grant 653884.

\bibliographystyle{IEEEtran}
\bibliography{IEEEabrv,lit_abbr}

\end{document}

%% file: abstraction_layers.pdf_tex
\begingroup%
  \makeatletter%
  \providecommand\color[2][]{%
    \errmessage{(Inkscape) Color is used for the text in Inkscape, but the package 'color.sty' is not loaded}%
    \renewcommand\color[2][]{}%
  }%
  \providecommand\transparent[1]{%
    \errmessage{(Inkscape) Transparency is used (non-zero) for the text in Inkscape, but the package 'transparent.sty' is not loaded}%
    \renewcommand\transparent[1]{}%
  }%
  \providecommand\rotatebox[2]{#2}%
  \ifx\svgwidth\undefined%
    \setlength{\unitlength}{314.46933594bp}%
    \ifx\svgscale\undefined%
      \relax%
    \else%
      \setlength{\unitlength}{\unitlength * \real{\svgscale}}%
    \fi%
  \else%
    \setlength{\unitlength}{\svgwidth}%
  \fi%
  \global\let\svgwidth\undefined%
  \global\let\svgscale\undefined%
  \makeatother%
  \begin{picture}(1,0.81461825)%
    \put(0,0){\includegraphics[width=\unitlength]{abstraction_layers.pdf}}%
    \put(0.41529333,0.81349089){\color[rgb]{0,0,0}\makebox(0,0)[lt]{\begin{minipage}{0.49116754\unitlength}\raggedright Security Invariants\end{minipage}}}%
    \put(0.35885678,0.67903266){\color[rgb]{0,0,0}\makebox(0,0)[lt]{\begin{minipage}{1.05029543\unitlength}\raggedright Access Control Abstraction\end{minipage}}}%
    \put(0.40110339,0.34061351){\color[rgb]{0,0,0}\makebox(0,0)[lt]{\begin{minipage}{0.87948622\unitlength}\raggedright Interface Abstraction\end{minipage}}}%
    \put(0.44380088,0.00744507){\color[rgb]{0,0,0}\makebox(0,0)[lb]{\smash{Box Semantics}}}%
    \put(0.55742585,0.75065274){\color[rgb]{0,0,0}\makebox(0,0)[lt]{\begin{minipage}{0.22859745\unitlength}\raggedright \AccessControlBISinvars{}\end{minipage}}}%
    \put(0.53286383,0.57872082){\color[rgb]{0,0,0}\makebox(0,0)[lt]{\begin{minipage}{0.15518206\unitlength}\raggedright \InterfaceAbstractionMAPSAccessControl{}\end{minipage}}}%
    \put(0.44059411,0.23746362){\color[rgb]{0,0,0}\makebox(0,0)[lt]{\begin{minipage}{0.16281397\unitlength}\raggedright \BoxSemanticsMAPSInterfaceAbstraction{}\end{minipage}}}%
    \put(0.18131967,0.09336253){\color[rgb]{0,0,0}\rotatebox{90}{\makebox(0,0)[lt]{\begin{minipage}{0.46692801\unitlength}\raggedleft \boxSemanticsToAccessControlAbstraction{}\end{minipage}}}}%
    \put(0.24030749,0.61142556){\color[rgb]{0,0,0}\makebox(0,0)[lt]{\begin{minipage}{0.26903691\unitlength}\raggedright \InterfaceAbstractionToAccessControl{}\end{minipage}}}%
    \put(0.71034686,0.61774949){\color[rgb]{0,0,0}\makebox(0,0)[lt]{\begin{minipage}{0.20478944\unitlength}\raggedleft \AccessControlToInterfaceAbstraction{}\end{minipage}}}%
    \put(0.03052762,0.26649202){\color[rgb]{0,0,0}\makebox(0,0)[lt]{\begin{minipage}{0.31545206\unitlength}\raggedright translates\end{minipage}}}%
    \put(0.95618745,0.86378978){\color[rgb]{0,0,0}\rotatebox{-90}{\makebox(0,0)[lt]{\begin{minipage}{0.88472138\unitlength}\raggedleft \AccessControlToBoxSemantics{}\end{minipage}}}}%
    \put(0.03052762,0.05612248){\color[rgb]{0,0,0}\makebox(0,0)[lt]{\begin{minipage}{0.09374292\unitlength}\raggedright verifies\end{minipage}}}%
    \put(0.03013076,0.14678951){\color[rgb]{0,0,0}\makebox(0,0)[lt]{\begin{minipage}{0.07260634\unitlength}\raggedright maps\end{minipage}}}%
    \put(0.66695576,0.28674014){\color[rgb]{0,0,0}\makebox(0,0)[lt]{\begin{minipage}{0.24549293\unitlength}\raggedleft \InterfaceAbstractionToBoxSemantics{}\end{minipage}}}%
    \put(0.18409278,0.16057971){\color[rgb]{0,0,0}\makebox(0,0)[lt]{\begin{minipage}{0.19157907\unitlength}\raggedleft \BoxSemantics{}\end{minipage}}}%
  \end{picture}%
\endgroup%